\documentclass{sigchi}

\toappear{\scriptsize Permission to make digital or hard copies of all or part of this work for personal or classroom use is granted without fee provided that copies are not made or distributed for profit or commercial advantage and that copies bear this notice and the full citation on the first page. Copyrights for components of this work owned by others than the author(s) must be honored. Abstracting with credit is permitted. To copy otherwise, or republish, to post on servers or to redistribute to lists, requires prior specific permission and/or a fee. Request permissions from permissions@acm.org. \\
{\emph{CHI '20, April 25--30, 2020, Honolulu, HI, USA.} } \\
\copyright~2020 Association for Computing Machinery. \\
ACM ISBN 978-1-4503-6708-0/20/04\ ...\$15.00. \\
http://dx.doi.org/10.1145/3313831.3376590}

\usepackage{balance}       
\usepackage{graphics}      
\usepackage[T1]{fontenc}   
\usepackage{txfonts}
\usepackage{mathptmx}
\usepackage[pdflang={en-US},pdftex]{hyperref}
\usepackage{color}
\usepackage{booktabs}
\usepackage{textcomp}
\usepackage{multirow}
\usepackage{array,booktabs}
\usepackage{microtype}        
\usepackage{ccicons}          

\usepackage{todonotes}

\def\plaintitle{Questioning the AI: Informing Design Practices for Explainable AI User Experiences}

\def\emptyauthor{}
\def\plainkeywords{Authors' choice; of terms; separated; by
  semicolons; include commas, within terms only; this section is required.}

\makeatletter
\def\url@leostyle{%
  \@ifundefined{selectfont}{
    \def\UrlFont{\sf}
  }{
    \def\UrlFont{\small\bf\ttfamily}
  }}
\makeatother
\def\pprw{8.5in}
\def\pprh{11in}

\definecolor{linkColor}{RGB}{6,125,233}
\hypersetup{%
  pdftitle={\plaintitle},
  pdfauthor={\emptyauthor},
  pdfkeywords={\plainkeywords},
  pdfdisplaydoctitle=true, 
  bookmarksnumbered,
  pdfstartview={FitH},
  colorlinks,
  citecolor=black,
  filecolor=black,
  linkcolor=black,
  urlcolor=linkColor,
  breaklinks=true,
  hypertexnames=false
}

\begin{document}

\title{\plaintitle}

\numberofauthors{3}
\author{
  \alignauthor{Q. Vera Liao\\
   \affaddr{IBM Research AI}\\
    \affaddr{Yorktown Heights, NY, USA}\\
   \email{vera.liao@ibm.com}}\\
  \alignauthor{Daniel Gruen\\
    \affaddr{IBM Research}\\
    \affaddr{Cambridge, MA, USA}\\
    \email{daniel\_gruen@us.ibm.com}}\\
  \alignauthor{Sarah Miller\\
   \affaddr{IBM Research}\\
    \affaddr{Cambridge, MA, USA}\\
    \email{millers@us.ibm.com}}\\
}

\maketitle

\begin{abstract}
A surge of interest in explainable AI (XAI) has led to a vast collection of algorithmic work on the topic. While many recognize the necessity to incorporate explainability features in AI systems, how to address real-world user needs for understanding AI remains an open question. By interviewing 20 UX and design practitioners working on various AI products, we seek to identify gaps between the current XAI algorithmic work and practices to create explainable AI products. To do so, we develop an algorithm-informed \textit{XAI question bank} in which user needs for explainability are represented as prototypical questions users might ask about the AI, and use it as a study probe.  Our work contributes insights into the design space of XAI, informs efforts to support design practices in this space, and identifies opportunities for future XAI work. We also provide an extended XAI question bank and discuss how it can be used for creating user-centered XAI.
\end{abstract}






\keywords{Explainable AI; human-AI interaction; User experience}

\printccsdesc

\section{Introduction}
The rapidly growing adoption of Artificial Intelligence (AI), and Machine Learning (ML) technologies using opaque deep neural networks in particular, has spurred great academic and public interest in explainability to make AI algorithms understandable by people. This issue appears in popular press, industry practices~\cite{h2o,arya2019one}, regulations~\cite{gdpr}, as well as hundreds of recent papers published in AI and related disciplines. These XAI works often express an algorithm-centric view, relying on ``\textit{researchers' intuition of what constitutes a `good' explanation}''~\cite{miller2018explanation}. This is problematic because AI explanations are often demanded by lay users, who may not have deep technical understanding of AI, but hold preconception of what constitutes useful explanations for decisions made in a familiar domain. As an example, one of the most popular approaches to explain a prediction made by a ML classifier, as dozens of XAI algorithms strive to do~\cite{guidotti2019survey}, is by listing the features with the highest weights contributing to a model's prediction. For example, a model predicting a patient having the flu may explain by saying ``the symptoms of sneeze and headache are contributing to this prediction''~\cite{ribeiro2016should}. However, it is questionable whether such an explanation satisfies a doctor's needs to understand the AI, or adds significant value to a clinical decision-support tool.

To close the gap between XAI algorithms and user needs for effective transparency, the HCI community has called for interdisciplinary collaboration~\cite{abdul2018trends} and user-centered approaches to explainability~\cite{wang2019designing}.  This emerging area of work tends to either build on frameworks of human explanations from social science, or empirically study how explanation features impact user interaction with AI. In this paper, we take a complementary approach by investigating challenges faced by industry practitioners to create explainable AI products, with the goal of identifying gaps between the algorithmic work of XAI and what is needed to address real-world user needs. 

Recently, an increasing number of open-source toolkits (e.g.~\cite{dalex, h2o, alibi,arya2019one}) are making XAI techniques, which produce various forms of explanation for ``black-box'' ML models, accessible to practitioners. However, little is known about how to put these techniques from research literature into practice. As we will show, it is challenging work to bridge user needs and technical capabilities to create effective explainabilty features in AI products. This kind of work often falls to those with a bridging role in product teams--the design and user experience (UX) practitioners, whose job involves identifying user needs, communicating with developers and stakeholders, and creating design solutions based on demands and constraints on both sides.  We study, therefore, how AI explainability is approached by design and UX practitioners, explore together with them how XAI techniques can be applied in various products, and identify opportunities to better support their work and thus the creation of user-centered explainable AI applications. 
 
 Given the early status of XAI in industry practices, we anticipate a lack of established means to uncover user needs or a shared technical understanding. Therefore, we develop a novel probe to ground our investigation, namely an XAI algorithm informed \textit{question bank}. As an explanation can be seen as an answer to a question~\cite{miller2018explanation}, we represent user needs for explainability in terms of the questions a user might ask about the AI. Drawn on relevant ML literature and prior work on question-driven explanations in various domains, we create a list of prototypical user questions that can be addressed by current XAI algorithms. These questions thus represent the current availability of algorithmic methods for AI explainability, allowing us to explore how they can be applied in various AI products, and identify their limitations for addressing real-world user needs. Our contributions are threefold:
  \vspace{-0.3em}

\begin{itemize}
    \item We provide insights into how user needs for different types of explainability are presented in various AI products. We suggest how these user needs should be understood, prioritized and addressed. We also identify opportunities for future XAI work to better satisfy these user needs.
     \vspace{-0.4em}
    \item We summarize current challenges faced by design practitioners to create explainable AI products, including variability of user needs for explainability, discrepancies between algorithmic explanations and human explanations and a lack of support for design practices.
     \vspace{-0.4em}
    \item We present an extended XAI question bank (Figure~\ref{fig:question}) by combining algorithm-informed questions and user questions identified in the study. We discuss how it can be used as guidance and tool to support the needs specification work to create user-centered XAI applications.
    \vspace{-0.2em}
\end{itemize}{}


\section{Background}

 \vspace{-0.2em}
\subsection{Explainable artificial intelligence (XAI)}


Although XAI first appeared in expert systems almost four decades ago~\cite{clancey1983epistemology,swartout1983xplain}, it is gaining widespread visibility as a field focusing on ML interpretability~\cite{carvalho2019machine}. The term explainability is used by the research community with varying scope. In much of the ML literature, XAI aims to make the reasons behind a ML model's decisions comprehensible to humans~\cite{guidotti2019survey,lipton2016mythos,ribeiro2016should}. In a broader view, explainability encompasses everything that makes ML models transparent and understandable, also including information about the data, performance, etc.~\cite{arya2019one,hohman2019gamut}. Our view aligns with the latter.

Recent papers surveyed this rapidly growing field and identified its key research threads~\cite{adadi2018peeking, carvalho2019machine,gilpin2018explaining, guidotti2019survey, mohseni2018survey,ras2018explanation}. We will discuss taxonomies of XAI techniques in the next section. Another core thread is the evaluation of explanations, which answers whether an explanation is good enough, and how to compare different explanations. These questions are not only critical for choosing appropriate XAI techniques, but also underlie the development of intelligent systems that optimize the choice of explanation, such as interactive or personalized explanations~\cite{abdul2018trends,schneider2019personalized, weld2018challenge}. Toward this goal, many sought to define the desiderata of XAI~\cite{carvalho2019machine,guidotti2019survey,hohman2019gamut,robnik2018perturbation}, including fidelity, completeness, robustness, etc. Despite the conceptual discussions, there are few established means of quantifying explainability. Partly, the reason is that the effectiveness of an explanation is relative to the recipient, and on a philosophical ground, the question asked~\cite{bromberger1992we}. So the same explanation may be seen as more or less comprehensible to different users, or even to the same user engaged in a different understanding. Many therefore advocate that the evaluation of XAI needs to involve real users within the targeted application~\cite{doshi2017towards,hoffman2018metrics}. 

Given its recipient-dependent nature, it is clear that work on XAI must take a human-centered approach. By conducting a literature review in social science on how humans give and receive explanations, Miller identified a list of human-friendly characteristics of explanation that are not given sufficient attention in the algorithmic work of XAI, including contrastiveness (to a counterfactual case), selectivity, social process, focusing on the abnormal, etc. Wang et al.~\cite{wang2019designing} proposed a conceptual framework to connect XAI techniques and cognitive patterns in human-decision making to guide the design of XAI systems.


With a fundamental interest in creating user-centered technologies, the HCI community is seeing burgeoning efforts around designing and studying user interactions with explainable AI~\cite{binns2018s,cai2019effects,cheng2019explaining, dodge2019explaining,hohman2019gamut,kocielnik2019will,lai2018human,rader2018explanations}. As a literature analysis performed by Abdul et al.~\cite{abdul2018trends} shows, before this wave of work on ML systems, the HCI community have studied explainable systems in various contexts, most notably context-aware applications~\cite{bellotti2001intelligibility,lim2009assessing,lim2010toolkit}, recommender systems~\cite{herlocker2000explaining}, debugging tools~\cite{kulesza2015principles} and algorithmic transparency ~\cite{diakopoulos2015algorithmic,sandvig2014auditing}. Specific to XAI, recent studies largely focused on empirically understanding the effect of explanation features on users' interaction with and perception of ML systems,  usually through controlled lab or field experiments. Notably, although explanations were found to improve user understanding of the AI systems,  conclusions about its benefits for user trust and acceptance were mixed~\cite{cheng2019explaining,kocielnik2019will,lai2018human,poursabzi2018manipulating}, suggesting potential gaps between algorithmic explanations and end user needs.

Our work shares the goal of bridging between the algorithm-centric XAI and user-centered explanations. In contrast to prior work centered around end users, we focus on the people that engage in this bridging work, namely UX and design practitioners. By studying their current practices, we explore the largely undefined design space of XAI and identify challenges for creating explainable AI products.
 \vspace{-0.9em}
\subsection{Supporting AI practitioners}
We join a growing group of scholars studying the work of industry practitioners who create AI products~\cite{amershi2019guidelines,boukhelifa2017data,hohman2019gamut,holstein2019improving,muller2019data,rule2018exploration}. By better supporting their work, we can ameliorate downstream usability, ethical and societal problems associated with AI. For example, Boukhelifa et al.~\cite{boukhelifa2017data} interviewed 12 data scientists to understand their coping strategies around uncertainty in data science work, and proposed a process model for uncertainty-aware data analytics. Holstein et al. interviewed 35 ML practitioners to conduct the first investigation of commercial product teams' challenges for developing fairer ML systems, and identified the disconnect between their needs and the solutions proposed in the fair ML research literature.

Most studies of AI practitioners focused on data scientists. As creating explainable AI products requires a user-centered approach, design practitioners should also perform an indispensable role. Despite a growing body of HCI work on AI technologies, there is a lack of design guidelines for AI systems. One notable exception is a recent paper by Amershi et al.~\cite{amershi2019guidelines}, which synthesized a set of 18 usability guidelines for AI systems. Several of these guidelines (e.g., make clear what the system can do, how well it can do,  why it did what it did) are relevant to explainability, but they do not provide actionable guidance on how to actualize these capabilities. Meanwhile, recent papers explored design methods supporting the creation of explainable AI systems. Wolf~\cite{wolf2019explainability} proposed a scenario-based approach to identify user needs for explainability (``\textit{what people might need to
understand about AI systems}'') early on in system development. Eiband et al.~\cite{eiband2018bringing} proposed a stage-based participatory design process, which guides product-specific needs specification--\textit{what} to explain, followed by iterative design of solutions--\textit{how} to explain.

Our work is motivated by a similar pragmatic goal of supporting design practices of XAI. More specifically, in the face of increasingly available XAI techniques, we are interested in the design work that connects user needs and these technical capabilities. In particular, we probe the challenges to identify the suitability of XAI techniques. A recent stream of guidance in the public domain (e.g.~\cite{dalex,h2o,arya2019one}) on how to select among XAI algorithms suggest their suitability can be difficult to determine. More problematically, these guidelines are targeting data scientists, using criteria grounded in the development process (e.g., explaining data or features, pre- or post-training). They do not directly address end user needs for understanding AI, nor support the navigation of the design space of XAI. 

\vspace{-0.8em}
\subsection{Question-driven explanations}
Outside the ML field, many explored the space of user needs for explanation using a question-driven framework. Fundamentally, an explanation is ``\textit{an answer to a (why-) question}~\cite{miller2018explanation}.'' These questions are also user and context dependent, described as ``triggers'' by Hoffman et al.~\cite{hoffman2018metrics} representing ``\textit{tacitly an expression of a need for a certain kind of explanation...to satisfy certain user purposes of user goals}. ''

In the early generation of AI work, question-driven frameworks were used to generate explanations for knowledge-based systems~\cite{chandrasekaran1989explaining,gregor1999explanations,swartout1987making}. AQUA~\cite{ram1989question} is a reasoning model that uses questions to generate explanations and identify knowledge gaps for learning. AQUA was built upon a taxonomy of questions, including anomaly detection questions, hypothesis verification questions, etc. Silveira et al. provided a taxonomy of user questions about software to drive the design of help systems~\cite{silveira2001semiotic}. Building on it, Glass et al.~\cite{glass2008toward} investigated users' explanation requirements in using an adaptive agent and showed that user needs for different types of explanation vary. In context-aware computing, Lim and Dey~\cite{lim2009assessing} developed a taxonomy of user needs for intelligibility by crowdsourcing user questions in scenarios of context-aware applications. These questions were coded into \textit{intelligibility types}, including input, output, conceptual model (why, how, why not, what else, what if) and non-functional types (certainty, control). This taxonomy enabled a toolkit that supports the generation of explanations for context-aware applications~\cite{lim2010toolkit}.


Inspired by the prior work, we use an XAI question bank, containing prototypical questions that users may ask for understanding AI systems, as a study probe representing user needs for AI explanability. Instead of using question taxonomies that are not specific to ML, we start by performing a literature review to arrive at a taxonomy of existing XAI techniques, and use it to guide the creation of user questions. Thereby we constrain the probe to reflect the current availability of XAI techniques to understand how user needs for such explainability are presented in real-world AI products.

\vspace{-0.3em}
\section{XAI question bank}
We now describe how we developed the XAI question bank by first identifying a list of explanation methods supported by current XAI algorithms, for which we focus on those generating post-hoc explanations for opaque ML models~\cite{arrieta2019explainable,guidotti2019survey}. For the scope of this paper, we will leave out the technical details of the algorithms but provide references for interested readers. There have been many efforts to create taxonomies of XAI methods~\cite{adadi2018peeking,arrieta2019explainable, carvalho2019machine,gilpin2018explaining, guidotti2019survey,mohseni2018survey,molnar2018interpretable,ras2018explanation,samek2019towards}. Commonly, they differentiate between an explanation \textit{method}--a pattern or mechanism to explain ML models--and specific XAI algorithms. One type of explanation method can be generated by multiple algorithms, which may vary in performance or applicability to specific ML models. Common dimensions to categorize explanation methods include: 1) The scope of the explanation, i.e. whether to support understanding the entire model (\textit{global}) versus a single prediction (\textit{local}); 2) The complexity of the model to be explained; 3) The dependency on the model used, i.e., whether the technique applies to any ML model or to only one type~\cite{adadi2018peeking}; and 4) The stage of model development to apply the explanation~\cite{carvalho2019machine}.

Except for the first one, these dimensions are data scientist centric as they are concerned with the characteristics of the underlying model. For our purpose of mapping user questions, we seek a taxonomy that lists the forms of explanation as presented to users. For example, we disregard the complexity of the model or the explanation's applicability to specific models, but instead differentiate between methods that describe the model logic as \textit{rules}, \textit{decision trees} or \textit{feature importance}.  Also,  to identify user questions an explanation addresses, we believe it is sufficient to stay at the general mechanism, and ignore the specificity of the presentation such as textual or by visualization \cite{ribeiro2016should}. Guided by these principles, we found the taxonomy of \textit{explanators} in Guidotti et al.~\cite{guidotti2019survey} closest to our purpose. Using it as a starting point, we consulted other survey papers and iteratively consolidated a taxonomy of explanation methods. In addition to the three categories in~\cite{guidotti2019survey}--methods that explain the entire model (\textit{global}), an individual outcome (\textit{local}), and inspect how the output changes with instance changes (\textit{inspect counterfactual}), we added \textit{example based} explanations \cite{hohman2019gamut,molnar2018interpretable}, since they represent a distinct mechanism to explain. Finally, we arrived at the taxonomy presented in the second column of Table~\ref{tab:xai}.

\begin{table*}[t]
  \centering
    {\small
  \begin{tabular}{>{\raggedright}p{1.6cm}>{\raggedright}p{3.2cm}|p{6.6cm}|p{1.4cm}|p{3.3cm}}\toprule
  Category of Methods& Explanation Method& Definition& Algorithm Examples& Question Type\\
  \midrule
  
Explain the model & Global feature importance & Describe the weights of features used by the model (including visualization that shows the weights of features) & \cite{henelius2014peek,lou2013accurate,nguyen2016multifaceted,tolomei2017interpretable}&\textbf{How} \\[0.35cm]
(\textbf{Global})& Decision tree approximation& Approximate the model to an interpretable decision-tree& \cite{bastani2017interpretability,johansson2009evolving,krishnan1999extracting}& \textbf{How}, Why, Why not, What if\\[0.1cm]

  &Rule extraction& Approximate the model to a set of rules, e.g., if-then rules & \cite{dash2018boolean,wei2019generalized,zhou2003extracting}& \textbf{How}, Why, Why not, What if\\
  \cmidrule{1-5}

Explain a prediction  & Local feature importance and saliency method & Show how features of the instance contribute to the model's prediction (including causes in parts of an image or text) & \cite{lundberg2017unified,ribeiro2016should,simonyan2013deep,vstrumbelj2014explaining,zhou2016learning}& \textbf{Why}\\[0.4cm]

(\textbf{Local})& Local rules or trees& Describe the rules or a decision-tree path that the instance fits to guarantee the prediction &\cite{guidotti2018local,ribeiro2018anchors,zhang2019interpreting} & \textbf{Why}, \textbf{How to still be this} \\
  \cmidrule{1-5}
  
  \textbf{Inspect counterfactual} & Feature influence or relevance method &  Show how the prediction changes corresponding to changes of a feature (often in a visualization format) & \cite{apley2016visualizing,friedman2001greedy,goldstein2015peeking,krause2016interacting} & \textbf{What if}, How to be that, How to still be this\\[0.1cm]

 &   Contrastive or counterfactual features & Describe the feature(s) that will change the prediction if perturbed, absent or present   & \cite{dhurandhar2018explanations,wachter2017counterfactual,zhang2018interpreting}  & \textbf{Why}, \textbf{Why not}, \textbf{How to be that}\\
   \cmidrule{1-5}
  \textbf{Example based} & Prototypical or representative examples& Provide example(s) similar to the instance and with the same record as the prediction & \cite{bien2011prototype,kim2014bayesian,koh2017understanding}& \textbf{Why}, How to still be this\\[0.4cm]
 
  & Counterfactual example & Provide example(s) with small differences from the instance but with a different record from the prediction &\cite{goodfellow2014explaining,laugel2017inverse,mothilal2020explaining}& \textbf{Why}, \textbf{Why not}, How to be that\\

  \bottomrule
  
  \end{tabular}
  }
 \vspace{-0.3em}
  \caption{ Taxonomy of XAI methods mapping to user question types. Questions in bold are the primary ones that the XAI method addresses. Questions in regular font are ones that only a subset of cases the XAI method can address. For example, while a global decision tree approximation can potentially answer~\protect\textit{ Why, Why not, and What if} questions for individual instances~\protect\cite{lim2009and}, the approximation may not cover certain instances.}~\label{tab:xai}
  \vspace{-2em}
\end{table*}

To map the explanation methods to user questions they can answer, we consulted prior work that provided taxonomies of questions for explanations~\cite{lim2009assessing,lim2010toolkit, ram1989question,silveira2001semiotic}. The closest to our purpose is the \textit{intelligibility types} by Lim et al. ~\cite{lim2009assessing,lim2010toolkit}, developed by eliciting user questions in scenarios of context-aware computing. In particular, the intelligibility types of \textit{How} (system logic), \textit{Why} (a prediction), \textit{Why not}, \textit{What if} are directly applicable to ML systems. By mapping these questions to explanation methods listed in Table~\ref{tab:xai}, we identified two additional types of question that can be addressed by existing XAI techniques: 1) \textit{How to be that}: what are the changes required, often implying minimum changes, for an instance to get a different target prediction; 2) \textit{How to still be this}: what are the permitted changes, often implying maximum changes, for an instance to still get the same prediction. We note that the questions of \textit{What if}, \textit{How to be...} are considered \textit{counterfactual questions} and best answered by inspection or example based explanations, which allow users to understand the decision boundaries of a ML model. Table~\ref{tab:xai} was reviewed by 4 additional experts working in the field of XAI.

Taking a broad view on explainability, we also consider descriptive information that could make a ML model more transparent. We added three more types based on \cite{hohman2019gamut, lim2009assessing,lim2010toolkit}-- questions regarding model \textit{Input} (training data), \textit{Output}, and \textit{Performance}. In the rest of the paper, we refer these 9 types of questions as 9 \textit{explainability needs categories}  as they represent categories of prototypical questions users may ask to understand AI. For each category, we created a leading question (e.g.,``\textit{Why is this instance given this prediction}'' for the \textit{Why} category\footnote{\vspace{-0.3em}{\scriptsize We instructed that `prediction' is used to refer the AI output for an instance. In the context of a product, it can mean a score/ recommendation/ classification/ answer, etc.}}), and supplemented 2-3 additional example questions, inquiring about features and examples whenever applies. The list of questions developed in this step are shown in Figure~\ref{fig:question} without an asterisk. We do not claim the exhaustiveness of this list, but deem it to be sufficient as a study probe.
\vspace{-0.5em}
\section{Study design}
We conducted semi-structured interviews with 20 UX and design practitioners recruited from multiple product lines at IBM. All but two (I-6 and I-20) informants worked on different products without shared AI models. Three informants were design team leads overseeing multiple products. The AI products included mature commercial platforms, systems in the testing phase, and internal platforms used by IBM employees. 50\% of informants were female. All but two were based in the United States, in 7 different locations.  Table~\ref{tab:informants} summarizes the primary areas of the products and informants' job titles. Our samples focused on AI systems that support utility tasks such as analytics and decision-support, as explainability is critical for high-stakes tasks where people would want to understand the AI's decisions~\cite{carvalho2019machine,doshi2017towards}. Informants were recruited from internal chat groups relevant to design and UX of AI products. The recruiting criteria indicated that one should have worked on the design of an AI product and had a good understanding of its users, and mentioned that the interview would focus on user needs around understanding the AI. 

\begin{table}[t]
  \centering
    {\small
  \begin{tabular}
{>{\raggedright}p{2.8cm}>{\raggedright}p{3.2cm}p{1.5cm}}\toprule
Technology area & Job title & Informant IDs\\
\midrule
Business decision support & HCI researcher, Designer, Designer, Data scientist, User researcher & I-4, I-5, I-12, I-17, I-19\\

Medical analytic or decision support& Product lead, Design researcher, Design researcher, Designer, Design researcher& I-1, I-6, I-7, I-11, I-20\\

AI model training or customization tools & Designer, Project manager, Designer & I-10, I-14, I-15\\

Human resource support & Designer & I-3\\

Enterprise social & User researcher & I-9\\
Natural resource analytic & UX researcher & I-2\\
Customer care chatbot & UX researcher & I-13\\
Multiple areas & Design team leads & I-8, I-16, I-18\\
  \bottomrule
\end{tabular}
  }
 \vspace{-0.5em}
  \caption{Informants information}~\label{tab:informants}
  \vspace{-3.2em}
\end{table}

We noticed that the current status of explainability in commercial AI products vary--about two thirds of the products (68.8\%) have descriptive explanations about the data or algorithm, only a subset (37.5\%) provide explanations for individual decisions, and certain products (e.g., chatbot) have neither. To explore the design space of XAI, we were interested in user needs for explainability uncovered by the design practitioners instead of the current system capabilities. The XAI question bank was able to scaffold the discussions, both to enumerate on the explainability needs categories, and to ground the discussion on user questions instead of venturing into the technical details. 

Using MURAL--a visual collaboration tool, we created a card for each question category listed in Figure~\ref{fig:question}, with the leading and example questions (without an asterisk). Informant went through each card and discussed whether they encountered these questions from users; If not, we asked whether they saw the questions would apply and in what situations. After pilot testing, for efficiency, we combined the \textit{Why} and \textit{Why not} into one card to represent user needs to understand a prediction; and \textit{What if}, \textit{How to be that}, \textit{How to still be this} into one card to represent user needs to understand counterfactual cases.  Thus there were 6 cards plus a blank card if one wanted to create an additional category. If time permitted, we asked informants to sort the cards according to their priority to address, and elicited the reasons for the ordering. 

Interviews lasted 45-60 minutes, conducted remotely using a video conferencing system and MURAL. We started by asking informants to pick an AI product they worked on and had good knowledge of the users, in which they saw user needs for understanding the AI. We asked them to describe the system and the AI components. They could either use screen sharing or send screenshots to show us the system. We then asked whether the users had \textit{needs to understand the AI}, and probed on why, when and where they had such needs (or lack thereof), and how the needs could be addressed, currently or speculatively. We then asked informants to reflect on what questions users would ask about the AI and listed as many as they could. User questions were also added to MURAL by the researchers if they appeared in other parts of the discussion. Thereby, we gathered user questions in a bottom-up fashion that allowed us to identify gaps in the algorithm-informed XAI question bank. It also prepared informants to move to discussions around the question cards. We closed the interview by asking informants to reflect on common challenges to build explainability features in AI products, and what kind of support they wished to have. For the three informants on lead roles, we focused on discussing the general status of explainability in AI products. 

\vspace{-0.5em}
\subsection{Analysis}
Around 1000 minutes of interviews were recorded and transcribed, from which we extracted 607 passages broadly relevant to explainability. We performed open coding and axial coding on these passages as informed by Grounded Theory research~\cite{corbin2015basics}. The iterative coding was conducted by one researcher, with frequent discussions with the other researchers. We returned to the passages, interview video and the AI products repeatedly as necessary. The iterative coding process resulted in a set of 24 axial codes. We combined them into selective codes to be discussed as the main themes in the results section, where the axial codes are presented in bold. 

Two additional sets of code were applied: 1)We identified 170 \textit{user questions} appeared in the question-listing activity and the rest of the interviews. 2)We coded these questions and other passages, wherever applied, with the \textit{explainability needs category}. The intersection of the two sets of code was 124 \textit{covered questions}, as covered by the categories of the question bank, and the remaining 46 \textit{uncovered questions}. 

\begin{figure*}
  \centering
  \includegraphics[width=1.7\columnwidth]{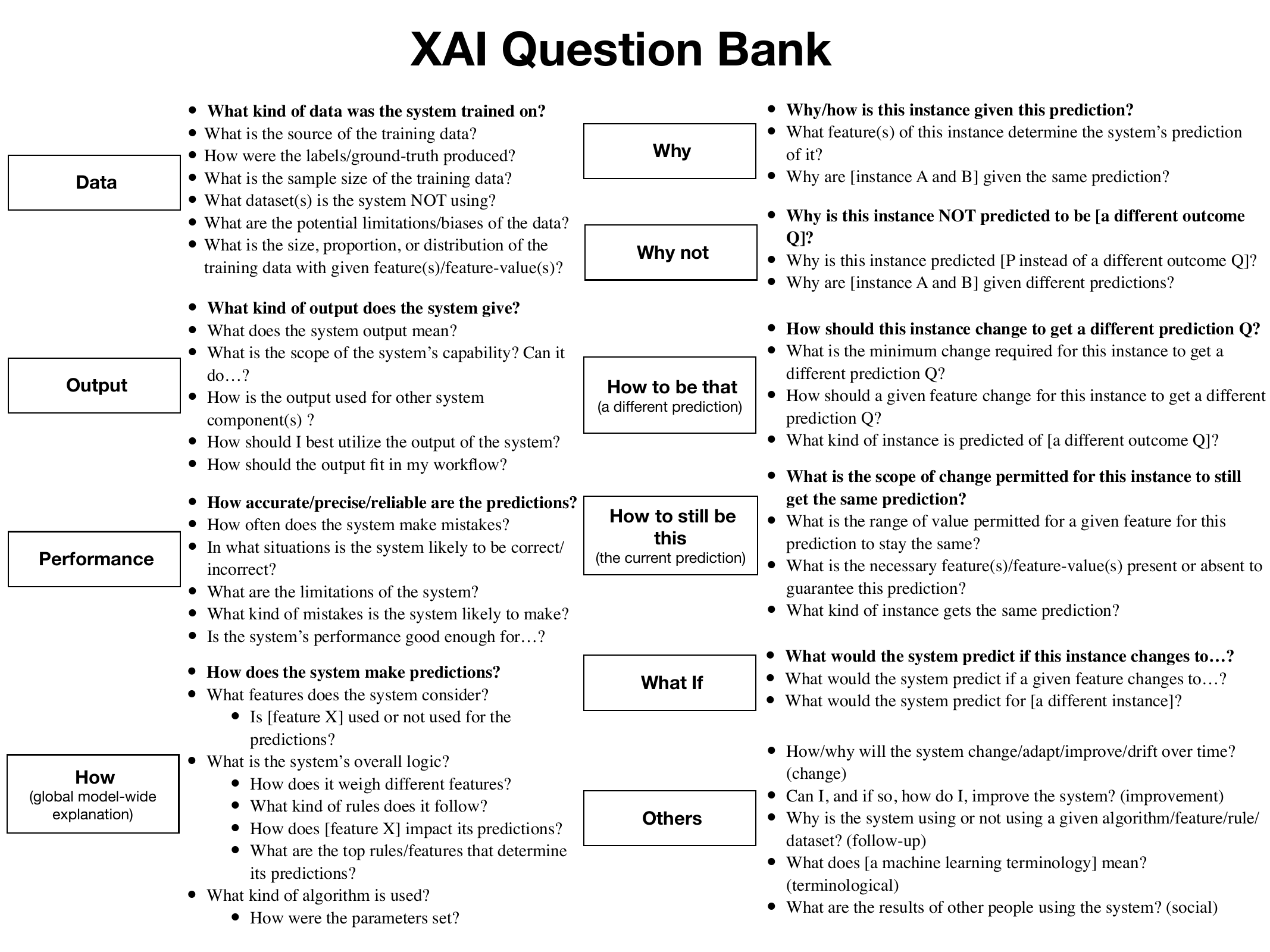}
   \vspace{-0.4em}
  \caption{XAI Question Bank, with leading questions in bold, and new questions identified from the interviews with \** (updated September 2021)}~\label{fig:question}
    \vspace{-2.6em}
\end{figure*}

To perform gap analysis on the XAI question bank, we followed two steps. For the covered questions in each needs category, we identified new forms of questions that were not covered by the original example questions, as shown with asterisks in Figure~\ref{fig:question}. By \textit{forms}, we grouped together questions with the same intent but phrased differently. For example, ``\textit{how was the data created}'', and ``\textit{where did the data come from}'' were both regarding the source of the training data, and covered by an original question in the \textit{Input} category, while ``\textit{what is the sample size}'' would be regarding a different characteristic of the input. In the second step, we examined the 46 uncovered questions. We first excluded 22 questions not generalizable to AI products, such as ``\textit{what is the summary of the article?}''. We then iteratively grouped and coded the intent of the remaining 24 questions and identified 5 additional forms of question in the \textit{Others} category in Figure~\ref{fig:question}. Insights from the analysis will be discussed the results section. 

\vspace{-0.5em}
\section{Results}
The results are divided into two parts. We start by discussing the general themes emerged in the interviews around the design work to create explainable AI products, which highlight some of the gaps between the algorithmic perspective of XAI and the practices to address user needs to understand AI. We then discuss how each category of user needs for explainability is presented in real-world AI products and based on that, reflect on the opportunities and limitations of XAI work.

\vspace{-0.5em}

\subsection{From XAI algorithms to design practices}

\subsubsection{The diverse motivations for and utility of explainability}

The historical context for the surge of XAI can be attributed to a fear of lacking understanding and control on increasingly complex ML models. Explanation is often embraced as a cure for ``black box'' models to gain user trust and adoption.  So a common pursuit is to produce interpretable, often simplified, descriptions of model logic to make an opaque ML model seen as transparent. This is a necessary effort, but insufficient to deliver a satisfying user experience if we ignore users' motivation for explanations. As I-8 put: ``\textit{Explainability isn't just telling me how you get there,  but also, can you expand on what you just told me...explanation has its own utility}''.

We identified several utility goals driving user demands for explanations of AI. In the context of AI-assisted decision-making, explanations are most frequently sought to \textbf{gain further insights or evidence}, as users are not satisfied by merely seeing a recommendation or score given by the AI. There are several ways people use these insights. When seeing disagreeable, unexpected or unfamiliar output, explanations are critical for people to assess the AI's judgment to \textbf{make an informed decision}. Even when users' decision aligns with the AI's, explanations could help \textbf{enhance decision confidence} or \textbf{generate hypothesis about the causality} for follow-up actions, as illustrated by I-5's comment, who worked on a tool supporting supply chain management:  ``\textit{users need to know why the system is saying this will be late because the reason is going to determine what their next action is...If it's because of a weather event, so no matter what you do you're not going to improve this number, versus something small,if you just make a quick call, you can get that number down}.'' In some cases, users also deem explanations of the AI's decision as potential \textbf{mitigation of their own decision biases}.

To \textbf{appropriately evaluate the capability of} the AI system is identified as the second theme of motivation, both to \textbf{determine the overall system adoption}  (e.g., evaluating data quality, transferability to a new context, whether the model logic aligns with domain knowledge), and at the operational level to \textbf{beware of the system's limitations}. I-6 commented on why explanations matter for users of a medical imaging system: ``\textit{There is a calibration of trust, whether people will use it over time.  But also saying hey, we know this fails in this way}.'' We note that appropriating trust should be distinguished from \textit{enhancing trust}. Though from a product team's perspective, the concern is often on users' under-trusting of the AI system and explanations are sought to improve adoption.

The third theme of motivation for explainability is to \textbf{adapt usage or interaction behaviors to better utilize the AI}. I-7 described users' desire to understand how the AI extracted information from clinic notes so they could adapt their notes-taking practices. I-17 mentioned users of a sales inventory management tool would want to focus on cases where the AI prediction was likely to err. I-13 commented that explanation could suggest to chatbot users what kind of things they could ask. Furthermore, explanations could also \textbf{convince users to invest in the system} ``\textit{if they know how the system will improve}''(I-11) (e.g., access to personal information, feedback).

Several informants working on AI systems supporting analysts' work or model training tools considered explanations as an integral part of a ``\textit{feedback loop}'' (I-11) to \textbf{improve AI performance}. Such needs are not only seen in debugging tools~\cite{hohman2019gamut}, but also in cases where the user could manipulate the data or correct the instance: ``\textit{[Explaining] why it thinks we are where we are and the opportunity to say, `no, I need you to just understand that we're in Phase 2'''} (I-5).

Last but not least, informants reflected on their \textbf{ethical responsibilities} to provide explanations:``
    \textit{What are we responsible for as creators of tools... whether it's out of the kindness of our hearts or whether it's because there's a true risk to others or society... We have to provide that level of explainability}''(I-8).

While some of these motivations have been discussed conceptually in prior work~\cite{adadi2018peeking,arrieta2019explainable,carvalho2019machine,doshi2017towards,guidotti2019survey}, our study provides concrete examples in real-world AI products. It is worth noting that the motivation for explainability is grounded in users' system goals such as improving decision-making, so explanation is not merely to provide transparency but support downstream user actions such as adapting interactions or acting on the decisions. Unpacking the motivation and downstream user actions should ultimately guide the selection of explanation methods. For example, if the goal is to gain further insights, example-based explanations could be more useful than feature-based explanations that describe the algorithm's logic, as I-2 described: ``\textit{[Users of natural resource analytic tools] already rely a lot on the analogy... [Similar examples] are very good to their study, e.g. give clues on which year this was formed.}''

Moreover, unpacking the motivation may help foresee the limitations of algorithmic explanations and fill the gaps in designing user experiences. For example, if the motivation is to mitigate biases, then users may desire to see ``\textit{both positive and negative evidence}'' (I-1). If it is to support adaption of interaction, the system could supplement information of ``\textit{this is what other people do}'' (I-13). Some informants criticized designing for the mental model of AI as a decision-maker explaining its rationale, and argued to focus on what utility explanations could provide to support users' end goals. As I-1, who worked on a clinical decision-support tool, put:``\textit{[explanations by system rationale] are essentially `this is how I do it, take it or leave it'. But doctors don't like this approach...Thinking that [AI is] giving treatment recommendations is the wrong place to start, because doctors know how to do it. It's everything that happens around that decision they need help with... more discussions about the output, rather than how you get there.}''

\vspace{-0.5em}

\subsubsection{In quest for human-like explanations}
Explanation is an integral part of human communication, and invites preconception of what constitutes effective and natural ways to explain. Informants were constantly dealing with discrepancies between algorithmic and human explanations. Some of the discrepancies are inherent to the mechanisms of AI algorithms, such as using features or learned patterns that are not aligned with how people make decisions. Others are in the forms constrained by technical capabilities and foreign to human interactions. For example: ``\textit{People have this unspoken norm, I trust you and if you are not sure you would let me know.  But nobody goes around saying how confident they are in the thing that they're saying. It may be implied in the language they're using.  So a system has high precision but 67\% confidence...it is a stupid and hard to use metric
}'' (I-5).

Several informants attempted to \textbf{mimic how people, especially domain experts, explain} in their design work. By aligning with how humans explain, it aligns user perception of the AI with existing mental model of decision-making in the domain, suggested by prior work as critical to build trust~\cite{springer2019progressive}. This is best exemplified by I-1's work in designing explanations for a clinical-decision support system that performs information extraction from medical literature: ``\textit{We mirror the way a doctor would do. So if a doctor was asked, how would you go and find the evidence?  ... You went to PubMed you found a paper, the paper matches my patient... you're showing me the statements in the paper [on whether] it was a good or bad idea, and putting all that together...So when you manage to reflect with AI literally how the doctor thinks about the problem, the black box kind of disappears.}''

We identified several themes on the desirable properties of human explanation that echoed Miller's arguments on informing XAI with how humans explain~\cite{miller2018explanation}. First, explanations are \textbf{selected}, often focusing on one or two causes from a sometimes infinite number of causes~\cite{miller2018explanation}.  Informants discussed the importance of selectivity as ``\textit{a balance of providing enough information that is trustworthy and compelling without overwhelming the user}'' (I-11) and acknowledging that ``\textit{AI will have a degree of [randomness] and may not be 100\% explainable}'' (I-8). Second, explanations are \textbf{social}, as part of a conversation and presented based on the recipient's current beliefs~\cite{miller2018explanation}. This social aspect is not only seen in tailoring explanations ``\textit{for people with different backgrounds}'' (I-4), but also in accommodating the evolving needs for explanation as one builds understanding and trust during the interaction process, ``\textit{once we trust it, it's about going deeper into it, the kind of questions goes from broad to ultra-specific }'' (I-8).

The selective and social nature of explanation has made many to argue that XAI has to be \textit{interactive} or even \textit{conversational}~\cite{madumal2019grounded,miller2018explanation,weld2018challenge}, tailoring explanation to different questions asked by different users, who would also ask follow-up questions to keep closing the gap of understanding, a process known as grounding in human communication~\cite{clark1991grounding}. Following prior work ~\cite{lim2009assessing,lim2010toolkit,ram1989question, weld2018challenge}, we postulate that a question-driven framework provides a viable path to interactive explanations.

\vspace{-0.6em}
\subsubsection{XAI: challenges and needs in design practices}
It is challenging work to create design solutions that bridge user needs for explainability and technical capabilities. The satisfaction of user needs is frequently \textbf{hampered by the current availability of XAI techniques}, which we will discuss in detail in the next section. Informants also had to work with other \textbf{product goals that are at odds with explainability}, such as protecting proprietary data, avoiding legal or marketing concerns from exposing the AI algorithm. Sometimes, explainability presents challenges to other aspects of user experience --``\textit{any opportunities we have to give them more explainability comes at the cost of the seamless integration.  And [doctors] are just so clear that not breaking their workflow is the most important factor to their happiness}''(I-6). Or, it might expose corner cases or rationales that some individuals found ``\textit{wrong}'' (I-2), ``\textit{over-simplified}'' (I-9), or ``\textit{outdated}'' (I-5), resulting in unnecessary user aversion and making the product ``\textit{victim of trying to be too transparent}'' (I-2).

In addition, unlike XAI algorithmic work's focus on one or a class of AI algorithm, creating explainable AI products requires \textbf{a holistic approach} that links multiple algorithms or system components to accommodate users' goal of better understanding and interacting with the system, as described by I-4: ``\textit{There is the traditional what we think about XAI, explaining what the model is doing. But there is this huge wrapper or the situation around it that people are really uncertain... what do I need to do with this output, how do I incorporate it into other processes, other tools?  So it is thinking about it as part of complex systems rather than one model and one output.}''

In short, inherent tension often exits between explainability and other system and business goals. Design practitioners are eager to advocate for explainability, but the realization requires teamwork with data scientists, developers and other stakeholders. Their advocacy is often hindered by \textbf{skill gaps} to engage themselves and the team in ``\textit{finding the right pairing to put the ideas of what's right for the user together with what's doable given the tools or the algorithms}''(I-8), and the \textbf{cost of time and resource} that a product team is reluctant to invest with a release schedule. These challenges can potentially be overcome by having resources that help sensitizing designers to the design opportunities in algorithmic explanations~\cite{yang2018machine} and enable conversations with the rest of the team, as expressed by many informants. We summarize informants' desirable support for XAI design in two areas:
\vspace{-0.3em}
\begin{enumerate}
    \item Guidance for \textbf{explainability needs specification}, for which we saw requests for both: 1) general principles of what types of explainability should be provided, as heuristic guidelines that a product can be developed or evaluated with; 2) guidance to identify product, user, and context specific needs to help the product team prioritize the effort.
      \vspace{-0.8em}
    \item Guidance for creating \textbf{explainability solutions} to address user needs, paired with example artifacts (e.g., UI elements, design patterns), to support the exploration of tangible solutions and communication with developers and stakeholders. 
\end{enumerate}{}
\vspace{-0.3em}
The two areas correspond to the \textit{what to explain} and \textit{how to explain} stages in Eiband et al.'s design process for transparent interfaces~\cite{eiband2018bringing}. We argue that the question bank could potentially support needs specification work, as it essentially lays out the space of users' prototypical questions to understand AI systems. The above requests suggest the needs to further understand the key factors that may lead to variability of user questions, and how these questions should be appropriately answered. We work towards these goals in the next section.



\vspace{-0.3em}

\subsection{Understanding user needs for explainability}
We use the \textit{explainability needs category} codes to guide our analysis on each category. We focus on two questions: 1) The variability of explainability needs, i.e., what factors make a category of user questions more or less likely to be asked. 2) The potential gaps between algorithmic explanations and user needs, by examining passages coded as design challenge, and the additional questions identified in the gap analysis (Figure~\ref{fig:question}). To help answer the former, we first discuss key factors that may lead to the variability of explainability needs, which we identified by coding informants' reasons to include, exclude or prioritize a needs category. 
\vspace{-0.3em}
\begin{itemize}
    \item \textbf{Motivation for explainability}:The diverse motivations discussed in the last section for demanding explainability could lead to wanting different kinds of explanation. 
     \vspace{-0.5em}
    \item \textbf{Usage point}: Informants mentioned common points during the usage of AI systems where certain type of explianabiltiy was of particular interest, including on-boarding, reliance or delegation to AI, encountering abnormal results, system breakdown, and seeing changes in the system.
     \vspace{-0.5em}
    \item \textbf{Algorithm or data type}: Different algorithms invoke different questions. For lay users, it might be more relevant to consider the type of data the AI is used with rather than specific algorithms, e.g., tabular data, text, images or video.
    \vspace{-1.3em}
    \item \textbf{Decision context}: We identified codes describing the nature of the decision context that led to prominent needs for certain type of explainability, including outcome criticality, time-sensitivity, and decision complexity.
     \vspace{-0.5em}
    \item \textbf{User type}: Codes describing the characteristics of users include AI knowledge, domain knowledge, attitude towards AI, roles or responsibilities.
    
\end{itemize}{}

In prior work, the variability of user needs for explaianability has been discussed regarding the roles of the users~\cite{arrieta2019explainable,arya2019one,hind2019explaining,samek2019towards,weller2017challenges}, e.g., regulators, model developers, decision-makers, consumers. The diverse criteria used by our informants suggest many other factors to consider for the suitability of XAI techniques. This paper does not conclude on how these factors vary user needs. Rather, they should be seen as \textit{sensitizing concepts} by Bowen's~\cite{bowen2006grounded} and Ribes's~\cite{ribes2017notes} definitions--``\textit{tell where to look but not what to see.}'' The sheer number of these factors highlight the challenge to pre-define users' explainability needs, vindicating the recent effort~\cite{eiband2018bringing,wolf2019explainability} to provide structured guidance to support empirically identifying application-specific user needs. Below we present informants' discussions on each category of explainability needs and highlight how these factors heighten the needs (in italic). 
 \vspace{-0.2em}
 
\subsubsection{Input/data}
Understanding training data for the AI model was most frequently seen to serve the motivation to \textit{appropriately evaluate AI capabilities} for use. It was considered a prominent need during the \textit{on-boarding} stage, and by both the \textit{decision-makers} and people in \textit{quality-control roles}. Explanations of data were also important in cases where the users could directly manipulate the data to either \textit{adapt the usage to better utilize the AI} or to \textit{improve the AI performance}.

Additional questions identified from the gap analysis indicate a desire to gauge the AI's limitations by inquiring about the sample size, potential biases, sampling of critical sub-groups, missing data, etc. Additional codes include to understand the system's compliance with regulations regarding data sampling, and transferability of the AI model: ``\textit{Not necessarily source,  but more conceptual like...[are we] making the solutions based on what occurred yesterday}'' (I-4). These patterns imply that users demand \textbf{comprehensive transparency of training data, especially the limitations}.
\vspace{-0.3em}
\subsubsection{Output}
While understanding the output is often an neglected aspect in algorithmic work of XAI, we saw frequent questions on it, indicating users' desire to understand the value of the AI system to \textit{appropriately evaluate the capability} and to \textit{better utilize the AI}, often in the \textit{on-boarding stage} or dealing with \textit{complex decisions}. Explaining output and explaining input/data were considered as ``\textit{static explanations}'' that more likely come up in the early stage of system usage, instead of frequent ``\textit{day-to-day, or transaction-to-transaction interactions}'' (I-8).

The most frequently asked questions were not regarding descriptive information of the algorithmic output, but at a high level, inquiring how to best utilize the output. We also identified two additional questions--``\textit{the scope of the capability}'', and ``\textit{how the output impacts other system components}.'' To address such user needs requires \textbf{contextualizing explaining the system's output in downstream system tasks and the users' overall workflow}.

\vspace{-0.3em}

\subsubsection{Performance}
To our surprise, the performance category was repeatedly ranked at the bottom, especially for users \textit{without AI background} and in decision contexts considered \textit{less critical}. There was a common hesitation among informants to present ML performance metrics such as accuracy, not only because a numerical value could be hard to interpret by lay users, but also there may be discrepancy between performances on the test data and the actual data, creating different ``experienced accuracy''~\cite{yin2019understanding} that might deter users. Some also believed that small differences in these metrics would not change how users interact: ``\textit{Technically that's great, but, it's still not a hundred... there's always going to be work that the users have to do to verify or double check}'' (I-4).


As many informants pointed out, and suggested by the additional questions, the goal of explaining performance should be to \textbf{help users understand the limitations of the AI, and make it actionable} as to answer ``\textit{Is the performance good enough for...}.'' There are constraints of technical capabilities. For example, confidence scores were repeatedly dismissed as not providing enough actionability --``\textit{[users] struggle to really understand,  does it mean it's going to do what I want it to do, or, can I trust it? }'' (I-15). Regarding the additional question on ``\textit{What kind of mistake}'', informants mentioned the precision-recall trade-off is a deliberately decided limitation that should be explained as it might change users' course of actions~\cite{kocielnik2019will} :``\textit{It's use case dependent... for the [doctors] if they miss a tumor, that's a life changing. So they have a very high tolerance for false positives} ''(I-7).

\vspace{-0.3em}
\subsubsection{How--global model}
Informants recognized the importance of providing global explanations on how the AI made decisions, both to help users \textit{appropriately evaluate the system capabilities}, and build a mental model to \textit{better interact with} or \textit{improve the system}. Such needs were prominent in cases where users were in a \textit{quality-control} role, or in a position able to \textit{adjust the model or the data-collection process}: ``\textit{The company really care about which of these attributes are the most important... then they will forward the manufacturer to include those in the data''} (I-17). Informants also agreed that users with \textit{AI or analytic background} were more likely to seek global explanations.


As Table~\ref{tab:xai} shows, to answer the \textit{How} question, XAI algorithms commonly employ ranked features, decision trees or rules. However, some informants were referring to high-level descriptions, such as ``\textit{I would just say keywords matching, it is intuitive, and it's been around}'' (I-3). Some were also concerned about fitting a complete \textit{How} explanation into the users' workflow: ``\textit{I can't imagine [doctors are] going into their workflow and be like, I'm so busy, let me read more about this AI. But, they would probably want some kind of confirmation about how it makes decisions}'' (I-11). So the design challenge is to identify the \textbf{appropriate level of details to explain the model globally}. This challenge is reflected in the question bank as well. While most XAI methods focus on answering ``\textit{What is the overall logic}'', we discovered that many questions were simply asking about the top features or whether certain feature was used, meanwhile a small set of questions by users with AI background were regarding the technical details of the model.


 \vspace{-0.3em}

\subsubsection{Why, Why not--local prediction}

Understanding a particular decision was often ranked at the top, and in user questions mentioned in all products. These questions were naturally raised after a surprising or abnormal event: ``\textit{For everyday interactions, most likely it's how did the system give me this answer? Not just any answer, but all of a sudden, here's this thing that I'm [not expecting] seeing}'' (I-8). This pattern is pointed out by Miller~\cite{miller2018explanation} as the \textit{contrastive} nature of human explanations, which are often implying \textit{Why not} the expected event.  We observed a shared \textbf{struggle with available technical solutions answering \textit{Why} but not \textit{Why not}}. Several informants working on text-based ML commented on the inadequacy of the common approach by highlighting keywords that contribute to the prediction:``\textit{even though we explained conceptually how it's working, it wouldn't be able to explain that error. So it would actually be counter-intuitive why it should make that error}'' (I-4). I-17 discussed the limitation of a state-of-the-art explanation algorithm, LIME~\cite{robnik2018perturbation}, which generates feature importance for ``black-box'' ML models.  She found the static explanation to be unsatisfying: ``\textit{LIME would say `it is boot cut which is why [it's not going to sell]', but would it be different if it was a skinny cut?}''

Many current XAI algorithms focus on the \textit{Why} question. We note that a challenge for algorithmic explanations is that the contrastive outcome is often not explicitly available to the model.  These observations again suggest the benefit of interactive explanations, allowing users to explicitly reference the contrastive outcome and asking follow-up \textit{What if} questions.


 \vspace{-0.4em}
\subsubsection{What if, How to be--inspecting counterfactual}
This category of explainability needs was not ranked high, and informants mentioned only 3 related user questions. Currently, these kinds of explanation are not widely adopted in commercial AI products. As prior work suggested, awareness of new types of explainability could change user demand~\cite{lim2009assessing}. In fact, informants recognized its \textbf{potential utility as system features to test different scenarios} for users to \textit{gain further insights} for the decision, and to \textbf{understand the boundary of system capabilities} to enable \textit{adapting interaction behaviors}. Informants also identified that such features align with how data scientists currently debug to \textit{improve ML models}. For example, I-4 was excited to consider how \textit{What if} explanations might support supply chain managers make decisions--``\textit{you can run different scenarios... the system can make an initial recommendation and then they can tweak it to see, the impact on the cost after that.}''  I-13 speculated that \textit{How to be that} explanations (how the chatbot would understand differently) could help chatbot users better phrase their queries. I-15 working on a tool for customizing entity recognition models commented that seeing how instance changes impact the output could help users debug the training data.

As seen in Table~\ref{tab:xai}, there is a growing collection of XAI techniques addressing the counterfactual questions. However, currently the feature influence methods are mostly used in data science tools~\cite{hohman2019gamut}. Contrastive feature and example based methods are relatively new areas of XAI work~\cite{dhurandhar2018explanations,wachter2017counterfactual, zhang2018interpreting}. Our results suggest their potentials as utility features in a broad spectrum of AI products. Future work should explore these potentials and sensitize practitioners to these possibilities.
\vspace{-0.3em}
\subsubsection{Additional explainability needs}
We also identified a set of questions that were not covered by the algorithm-informed needs categories. They point to additional areas of interest that users have for understanding AI. A critical area is to \textbf{understand the changes and adaption of AI}, both in explaining changes in the system, and how users can change the system. Other areas are 
\textbf{follow-up questions} by further inquiring why a certain feature or data is used, and \textbf{terminological questions} such as ``\textit{what do you mean by...}'', both of which may naturally emerge in an interactive explanation paradigm. Lastly, some users might be interested in knowing other people's experience with the system, suggesting a new mechanism for an AI system to provide \textbf{social explanations} with regard to other users' outcomes and actions. 
\vspace{-0.6em}
\section{Discussion}
With widespread calls for transparent and responsible AI, industry practitioners are eager to take up the ideas and solutions from the XAI literature. However, despite recent effort toward a scientific understanding of human-AI interaction~\cite{dodge2019explaining,narayanan2018humans,zhu2018explainable}, XAI research still struggles with a lack of understanding of real-world user needs for AI transparency, and by far little consideration of what practitioners need to create explainable AI products. Our study suggests the following directions both for algorithmic work to close the gaps addressing user needs, and design support to reduce technical and practical barriers to create user-friendly XAI products.
\vspace{-0.3em}
\begin{itemize}
    \item XAI research should direct its attention to techniques that address user needs, and we suggest a question-driven framework to embody these needs. Our results point to a few common questions and their desired answers that future work of XAI should explore, for example, \textit{How} question answered by multi-level details describing the algorithm, \textit{Why not} question referencing an expected alternative outcome, and \textit{How/Why will the system change}. Considering the coverage of user questions, especially common and new questions identified, could help the community move toward more human-centric effort. The question bank presented in this paper is just a starting point. Future work could continue building the repository by directly eliciting questions from end users of different types of AI systems~\cite{lim2009assessing}.
    \vspace{-0.4em}
    \item Practitioners struggle with the gaps between algorithmic output and creating human-consumable explanations. To close the gap requires inter-disciplinary work that studies how humans explain, and formalizes the patterns in algorithmic forms. Such a practice has already been engaged in interactive ML~\cite{amershi2014power,stumpf2007toward} and "socially guided machine learning"~\cite{thomaz2006transparency}. Prior work repeatedly pointed out that a prerequisite for explanations to be truly human-like is to be \textit{interactive}~\cite{madumal2019grounded,miller2018explanation,weld2018challenge}, because explanation is a grounding process where people incrementally close the belief gaps. Indeed, our study found that some user questions are closely connected with or followed by other questions. Future work could explore interaction protocols, for example through statistical modeling of how humans ask different explanation-seeking questions~\cite{madumal2019grounded}, to drive the flow of interactive or conversational explanations.  
    \vspace{-0.4em}
    \item Our study revealed the variability of user questions and its complex mechanisms, highlighting the challenge to identify product-specific user needs. While prior work attempted at top-down descriptions of needs of users in different roles~\cite{arrieta2019explainable,arya2019one,hind2019explaining,samek2019towards,weller2017challenges}, it may not be sufficient for design work that has to consider specific actions, usage points, models, etc. Recent HCI work on XAI encourages empirically identifying user needs with structured procedures~\cite{eiband2018bringing,wolf2019explainability}. We suggest several ways the XAI question bank can be used for needs specification. First as heuristic guidance, a product team could enumerate on whether each question category has been addressed and which should be prioritized. Second, it can be used in user research to scaffold the elicitation of user needs. For example, card-sorting exercises of the questions can be performed with users (adaptation may be required for specific AI applications). We invite practitioners to use, revise and expand the XAI question bank.
    \vspace{-0.4em}
    \item The technical barriers for designers and practitioners in general to navigate the space of XAI remains a primary challenge for product teams to optimize XAI user experiences. To support design work for ML, Yang suggested research opportunities to ``sensitize designers to the breadth of ML capabilities''~\cite{yang2018machine}. Informants also expressed strong desire for support of technical discussions with data scientists and stakeholders, as mitigating the friction is critical for the success of their advocacy for explainability. An opportunity for sensitizing support is to create concrete mapping between user questions and algorithmic capabilities, serving as a shared cognitive artifact between the designers and data scientists. One example, perhaps over-simplified, is the taxonomy of XAI methods we presented in Table~\ref{tab:xai}.  We may envision a \textit{question-driven design process}: by user research, a design practitioner identifies the primary type of user question as \textit{what to explain} (e.g. How to be that), and also the requirements to address the question as \textit{How to explain}. Table~\ref{tab:xai} then suggests candidate explanation method(s) to answer the question (e.g. contrastive features). Together with a data scientist, the team find the most suitable solution to implement from the list of suggested algorithms, then work toward closing the gaps between the algorithmic output and user requirements to answer the question. By suggesting conceptually this question-driven design process, we invite the research community to develop more fine-grained frameworks of XAI features (e.g., considering UI patterns) that connect user questions and technical capabilities. 
 
\end{itemize}{}

  \vspace{-0.5em}
\subsection{Limitations} 
First of all, the user questions were explored through design practitioners instead of end users, so we cannot claim this is a complete analysis of user needs for explainability. The results only reflect design practitioners' views. Future work could study other roles involved in AI product development (e.g., data scientists) to better understand the challenges to create XAI products. Our product samples focus on ones supporting high-stakes tasks, where needs for explainability might have been greater, and the current status of XAI more advanced.  We do not claim the completeness of the XAI methods discussed, especially as this is a fast advancing research field. Practitioners' increasing accessibility to XAI techniques may also change the demands and concerns expressed in the study.  Finally, our informants worked for the same organization. Although this is not uncommon for studies of practitioners~\cite{amershi2019guidelines,erickson2008assistance,muller2019data} and we recruited informants from diverse product lines and locations, we acknowledge that design practices may be different in other companies or organizations.
\vspace{-0.5em}
\section{Conclusion}
Although the research field of XAI is experiencing exponential growth, there is little shared practices of designing user-friendly explainable AI applications. We take the position that the suitability of explanations is \textit{question} dependent and requires an understanding of user questions for a specific AI application. We develop an XAI question bank to bridge the spaces of user needs for AI explainability and technical capabilities provided by XAI work. Using it as a study probe, we explored together with industry design practitioners the opportunities and challenges in putting XAI techniques into practice. We illustrated the great variability of user questions that may subject to many motivational, contextual and individual factors. We also identified gaps between current algorithmic solutions of XAI and what's needed to deliver satisfying user experiences, in the types of user questions to address and how they are addressed. We join many others in this field advocating a user-centered approach to XAI ~\cite{abdul2018trends,doshi2017towards,miller2018explanation,wang2019designing}. Our work suggests opportunities for the HCI and AI communities, as well as industry practitioners and academics, to work together to advance the field of XAI through translational work and shared knowledge repository that maps between user needs for explainability and XAI technical solutions.

\section{Acknowledgements}
We thank all our anonymous participants. We also thank Zahra Ashktorab, Rachel Bellamy, Amit Dhurandhar, Werner Geyer, Michael Hind, Stephanie Houde, David Millen, Michael Muller, Chenhao Tan, Richard Tomsett, Kush Varshney, Justin Weisz, Yunfeng Zhang, and anonymous CHI 2020 reviewers for their helpful feedback.

\balance{}

\bibliographystyle{SIGCHI-Reference-Format}
\bibliography{sample.bib}


\begin{thebibliography}{100}


\ifx \showCODEN    \undefined \def \showCODEN     #1{\unskip}     \fi
\ifx \showDOI      \undefined \def \showDOI       #1{{\tt DOI:}\penalty0{#1}\ }
  \fi
\ifx \showISBNx    \undefined \def \showISBNx     #1{\unskip}     \fi
\ifx \showISBNxiii \undefined \def \showISBNxiii  #1{\unskip}     \fi
\ifx \showISSN     \undefined \def \showISSN      #1{\unskip}     \fi
\ifx \showLCCN     \undefined \def \showLCCN      #1{\unskip}     \fi
\ifx \shownote     \undefined \def \shownote      #1{#1}          \fi
\ifx \showarticletitle \undefined \def \showarticletitle #1{#1}   \fi
\ifx \showURL      \undefined \def \showURL       #1{#1}          \fi

\bibitem{dalex}
 2018.
\newblock DALEX: Descriptive Machine Learning EXplanations.
\newblock   (2018).
\newblock
\newblock
\shownote{Accessed September 18, 2019 from \url{http://uc-r.github.io/dalex}.}


\bibitem{h2o}
 2018.
\newblock H2O Driverless AI.
\newblock   (2018).
\newblock
\newblock
\shownote{Accessed September 18, 2019 from
  \url{https://www.h2o.ai/products/h2o-driverless-ai/}.}


\bibitem{alibi}
 2019.
\newblock Alibi.
\newblock   (2019).
\newblock
\newblock
\shownote{Accessed September 18, 2019 from
  \url{https://github.com/SeldonIO/alibi}.}


\bibitem{abdul2018trends}
{Ashraf Abdul}, {Jo Vermeulen}, {Danding Wang}, {Brian~Y Lim}, {and} {Mohan
  Kankanhalli}. 2018.
\newblock \showarticletitle{Trends and trajectories for explainable,
  accountable and intelligible systems: An hci research agenda}. In {\em
  Proceedings of the 2018 CHI conference on human factors in computing
  systems}. ACM, 582.
\newblock


\bibitem{adadi2018peeking}
{Amina Adadi} {and} {Mohammed Berrada}. 2018.
\newblock \showarticletitle{Peeking inside the black-box: A survey on
  Explainable Artificial Intelligence (XAI)}.
\newblock {\em IEEE Access\/}  {6} (2018), 52138--52160.
\newblock


\bibitem{amershi2014power}
{Saleema Amershi}, {Maya Cakmak}, {William~Bradley Knox}, {and} {Todd Kulesza}.
  2014.
\newblock \showarticletitle{Power to the people: The role of humans in
  interactive machine learning}.
\newblock {\em AI Magazine\/} {35}, 4 (2014), 105--120.
\newblock


\bibitem{amershi2019guidelines}
{Saleema Amershi}, {Dan Weld}, {Mihaela Vorvoreanu}, {Adam Fourney}, {Besmira
  Nushi}, {Penny Collisson}, {Jina Suh}, {Shamsi Iqbal}, {Paul~N Bennett},
  {Kori Inkpen}, {and} {others}. 2019.
\newblock \showarticletitle{Guidelines for human-AI interaction}. In {\em
  Proceedings of the 2019 CHI Conference on Human Factors in Computing
  Systems}. ACM, 3.
\newblock


\bibitem{apley2016visualizing}
{Daniel~W Apley}. 2016.
\newblock \showarticletitle{Visualizing the effects of predictor variables in
  black box supervised learning models}.
\newblock {\em arXiv preprint arXiv:1612.08468\/} (2016).
\newblock


\bibitem{arrieta2019explainable}
{Alejandro~Barredo Arrieta}, {Natalia D{\'\i}az-Rodr{\'\i}guez}, {Javier
  Del~Ser}, {Adrien Bennetot}, {Siham Tabik}, {Alberto Barbado}, {Salvador
  Garc{\'\i}a}, {Sergio Gil-L{\'o}pez}, {Daniel Molina}, {Richard Benjamins},
  {and} {others}. 2019.
\newblock \showarticletitle{Explainable Artificial Intelligence (XAI):
  Concepts, Taxonomies, Opportunities and Challenges toward Responsible AI}.
\newblock {\em Information Fusion\/} (2019).
\newblock


\bibitem{arya2019one}
{Vijay Arya}, {Rachel~KE Bellamy}, {Pin-Yu Chen}, {Amit Dhurandhar}, {Michael
  Hind}, {Samuel~C Hoffman}, {Stephanie Houde}, {Q~Vera Liao}, {Ronny Luss},
  {Aleksandra Mojsilovi{\'c}}, {and} {others}. 2019.
\newblock \showarticletitle{One Explanation Does Not Fit All: A Toolkit and
  Taxonomy of AI Explainability Techniques}.
\newblock {\em arXiv preprint arXiv:1909.03012\/} (2019).
\newblock


\bibitem{bastani2017interpretability}
{Osbert Bastani}, {Carolyn Kim}, {and} {Hamsa Bastani}. 2017.
\newblock \showarticletitle{Interpretability via model extraction}.
\newblock {\em arXiv preprint arXiv:1706.09773\/} (2017).
\newblock


\bibitem{bellotti2001intelligibility}
{Victoria Bellotti} {and} {Keith Edwards}. 2001.
\newblock \showarticletitle{Intelligibility and accountability: human
  considerations in context-aware systems}.
\newblock {\em Human--Computer Interaction\/} {16}, 2-4 (2001), 193--212.
\newblock


\bibitem{bien2011prototype}
{Jacob Bien}, {Robert Tibshirani}, {and} {others}. 2011.
\newblock \showarticletitle{Prototype selection for interpretable
  classification}.
\newblock {\em The Annals of Applied Statistics\/} {5}, 4 (2011), 2403--2424.
\newblock


\bibitem{binns2018s}
{Reuben Binns}, {Max Van~Kleek}, {Michael Veale}, {Ulrik Lyngs}, {Jun Zhao},
  {and} {Nigel Shadbolt}. 2018.
\newblock \showarticletitle{'It's Reducing a Human Being to a Percentage':
  Perceptions of Justice in Algorithmic Decisions}. In {\em Proceedings of the
  2018 CHI Conference on Human Factors in Computing Systems}. ACM, 377.
\newblock


\bibitem{boukhelifa2017data}
{Nadia Boukhelifa}, {Marc-Emmanuel Perrin}, {Samuel Huron}, {and} {James
  Eagan}. 2017.
\newblock \showarticletitle{How data workers cope with uncertainty: A task
  characterisation study}. In {\em Proceedings of the 2017 CHI Conference on
  Human Factors in Computing Systems}. ACM, 3645--3656.
\newblock


\bibitem{bowen2006grounded}
{Glenn~A Bowen}. 2006.
\newblock \showarticletitle{Grounded theory and sensitizing concepts}.
\newblock {\em International journal of qualitative methods\/} {5}, 3 (2006),
  12--23.
\newblock


\bibitem{bromberger1992we}
{Sylvain Bromberger}. 1992.
\newblock {\em On what we know we don't know: Explanation, theory, linguistics,
  and how questions shape them}.
\newblock University of Chicago Press.
\newblock


\bibitem{cai2019effects}
{Carrie~J Cai}, {Jonas Jongejan}, {and} {Jess Holbrook}. 2019.
\newblock \showarticletitle{The effects of example-based explanations in a
  machine learning interface}. In {\em Proceedings of the 24th International
  Conference on Intelligent User Interfaces}. ACM, 258--262.
\newblock


\bibitem{carvalho2019machine}
{Diogo~V Carvalho}, {Eduardo~M Pereira}, {and} {Jaime~S Cardoso}. 2019.
\newblock \showarticletitle{Machine Learning Interpretability: A Survey on
  Methods and Metrics}.
\newblock {\em Electronics\/} {8}, 8 (2019), 832.
\newblock


\bibitem{chandrasekaran1989explaining}
{Bruce Chandrasekaran}, {Michael~C Tanner}, {and} {John~R Josephson}. 1989.
\newblock \showarticletitle{Explaining control strategies in problem solving}.
\newblock {\em IEEE Intelligent Systems\/} 1 (1989), 9--15.
\newblock


\bibitem{cheng2019explaining}
{Hao-Fei Cheng}, {Ruotong Wang}, {Zheng Zhang}, {Fiona O'Connell}, {Terrance
  Gray}, {F~Maxwell Harper}, {and} {Haiyi Zhu}. 2019.
\newblock \showarticletitle{Explaining Decision-Making Algorithms through UI:
  Strategies to Help Non-Expert Stakeholders}. In {\em Proceedings of the 2019
  CHI Conference on Human Factors in Computing Systems}. ACM, 559.
\newblock


\bibitem{clancey1983epistemology}
{William~J Clancey}. 1983.
\newblock \showarticletitle{The epistemology of a rule-based expert system—a
  framework for explanation}.
\newblock {\em Artificial intelligence\/} {20}, 3 (1983), 215--251.
\newblock


\bibitem{clark1991grounding}
{Herbert~H Clark}, {Susan~E Brennan}, {and} {others}. 1991.
\newblock \showarticletitle{Grounding in communication}.
\newblock {\em Perspectives on socially shared cognition\/} {13}, 1991 (1991),
  127--149.
\newblock


\bibitem{gdpr}
{European Commission}. 2016.
\newblock General Data Protection Regulation.
\newblock   (2016).
\newblock
\newblock
\shownote{Accessed September 18, 2019 from \url{https://eugdpr.org/}.}


\bibitem{corbin2015basics}
{Juliet Corbin}, {Anselm~L Strauss}, {and} {Anselm Strauss}. 2015.
\newblock {\em Basics of qualitative research}.
\newblock sage.
\newblock


\bibitem{dash2018boolean}
{Sanjeeb Dash}, {Oktay Gunluk}, {and} {Dennis Wei}. 2018.
\newblock \showarticletitle{Boolean decision rules via column generation}. In
  {\em Advances in Neural Information Processing Systems}. 4655--4665.
\newblock


\bibitem{dhurandhar2018explanations}
{Amit Dhurandhar}, {Pin-Yu Chen}, {Ronny Luss}, {Chun-Chen Tu}, {Paishun Ting},
  {Karthikeyan Shanmugam}, {and} {Payel Das}. 2018.
\newblock \showarticletitle{Explanations based on the missing: Towards
  contrastive explanations with pertinent negatives}. In {\em Advances in
  Neural Information Processing Systems}. 592--603.
\newblock


\bibitem{diakopoulos2015algorithmic}
{Nicholas Diakopoulos}. 2015.
\newblock \showarticletitle{Algorithmic accountability: Journalistic
  investigation of computational power structures}.
\newblock {\em Digital journalism\/} {3}, 3 (2015), 398--415.
\newblock


\bibitem{dodge2019explaining}
{Jonathan Dodge}, {Q~Vera Liao}, {Yunfeng Zhang}, {Rachel~KE Bellamy}, {and}
  {Casey Dugan}. 2019.
\newblock \showarticletitle{Explaining models: an empirical study of how
  explanations impact fairness judgment}. In {\em Proceedings of the 24th
  International Conference on Intelligent User Interfaces}. ACM, 275--285.
\newblock


\bibitem{doshi2017towards}
{Finale Doshi-Velez} {and} {Been Kim}. 2017.
\newblock \showarticletitle{Towards a rigorous science of interpretable machine
  learning}.
\newblock {\em arXiv preprint arXiv:1702.08608\/} (2017).
\newblock


\bibitem{eiband2018bringing}
{Malin Eiband}, {Hanna Schneider}, {Mark Bilandzic}, {Julian Fazekas-Con},
  {Mareike Haug}, {and} {Heinrich Hussmann}. 2018.
\newblock \showarticletitle{Bringing transparency design into practice}. In
  {\em 23rd International Conference on Intelligent User Interfaces}. ACM,
  211--223.
\newblock


\bibitem{erickson2008assistance}
{Thomas Erickson}, {Catalina~M Danis}, {Wendy~A Kellogg}, {and} {Mary~E
  Helander}. 2008.
\newblock \showarticletitle{Assistance: the work practices of human
  administrative assistants and their implications for it and organizations}.
  In {\em Proceedings of the 2008 ACM conference on Computer supported
  cooperative work}. ACM, 609--618.
\newblock


\bibitem{friedman2001greedy}
{Jerome~H Friedman}. 2001.
\newblock \showarticletitle{Greedy function approximation: a gradient boosting
  machine}.
\newblock {\em Annals of statistics\/} (2001), 1189--1232.
\newblock


\bibitem{gilpin2018explaining}
{Leilani~H Gilpin}, {David Bau}, {Ben~Z Yuan}, {Ayesha Bajwa}, {Michael
  Specter}, {and} {Lalana Kagal}. 2018.
\newblock \showarticletitle{Explaining explanations: An overview of
  interpretability of machine learning}. In {\em 2018 IEEE 5th International
  Conference on data science and advanced analytics (DSAA)}. IEEE, 80--89.
\newblock


\bibitem{glass2008toward}
{Alyssa Glass}, {Deborah~L McGuinness}, {and} {Michael Wolverton}. 2008.
\newblock \showarticletitle{Toward establishing trust in adaptive agents}. In
  {\em Proceedings of the 13th international conference on Intelligent user
  interfaces}. ACM, 227--236.
\newblock


\bibitem{goldstein2015peeking}
{Alex Goldstein}, {Adam Kapelner}, {Justin Bleich}, {and} {Emil Pitkin}. 2015.
\newblock \showarticletitle{Peeking inside the black box: Visualizing
  statistical learning with plots of individual conditional expectation}.
\newblock {\em Journal of Computational and Graphical Statistics\/} {24}, 1
  (2015), 44--65.
\newblock


\bibitem{goodfellow2014explaining}
{Ian~J Goodfellow}, {Jonathon Shlens}, {and} {Christian Szegedy}. 2014.
\newblock \showarticletitle{Explaining and harnessing adversarial examples}.
\newblock {\em arXiv preprint arXiv:1412.6572\/} (2014).
\newblock


\bibitem{gregor1999explanations}
{Shirley Gregor} {and} {Izak Benbasat}. 1999.
\newblock \showarticletitle{Explanations from intelligent systems: Theoretical
  foundations and implications for practice}.
\newblock {\em MIS quarterly\/} (1999), 497--530.
\newblock


\bibitem{guidotti2018local}
{Riccardo Guidotti}, {Anna Monreale}, {Salvatore Ruggieri}, {Dino Pedreschi},
  {Franco Turini}, {and} {Fosca Giannotti}. 2018.
\newblock \showarticletitle{Local rule-based explanations of black box decision
  systems}.
\newblock {\em arXiv preprint arXiv:1805.10820\/} (2018).
\newblock


\bibitem{guidotti2019survey}
{Riccardo Guidotti}, {Anna Monreale}, {Salvatore Ruggieri}, {Franco Turini},
  {Fosca Giannotti}, {and} {Dino Pedreschi}. 2019.
\newblock \showarticletitle{A survey of methods for explaining black box
  models}.
\newblock {\em ACM computing surveys (CSUR)\/} {51}, 5 (2019), 93.
\newblock


\bibitem{henelius2014peek}
{Andreas Henelius}, {Kai Puolam{\"a}ki}, {Henrik Bostr{\"o}m}, {Lars Asker},
  {and} {Panagiotis Papapetrou}. 2014.
\newblock \showarticletitle{A peek into the black box: exploring classifiers by
  randomization}.
\newblock {\em Data mining and knowledge discovery\/} {28}, 5-6 (2014),
  1503--1529.
\newblock


\bibitem{herlocker2000explaining}
{Jonathan~L Herlocker}, {Joseph~A Konstan}, {and} {John Riedl}. 2000.
\newblock \showarticletitle{Explaining collaborative filtering
  recommendations}. In {\em Proceedings of the 2000 ACM conference on Computer
  supported cooperative work}. ACM, 241--250.
\newblock


\bibitem{hind2019explaining}
{Michael Hind}. 2019.
\newblock \showarticletitle{Explaining explainable AI}.
\newblock {\em XRDS: Crossroads, The ACM Magazine for Students\/} {25}, 3
  (2019), 16--19.
\newblock


\bibitem{hoffman2018metrics}
{Robert~R Hoffman}, {Shane~T Mueller}, {Gary Klein}, {and} {Jordan Litman}.
  2018.
\newblock \showarticletitle{Metrics for explainable AI: Challenges and
  prospects}.
\newblock {\em arXiv preprint arXiv:1812.04608\/} (2018).
\newblock


\bibitem{hohman2019gamut}
{Fred Hohman}, {Andrew Head}, {Rich Caruana}, {Robert DeLine}, {and} {Steven~M
  Drucker}. 2019.
\newblock \showarticletitle{Gamut: A design probe to understand how data
  scientists understand machine learning models}. In {\em Proceedings of the
  2019 CHI Conference on Human Factors in Computing Systems}. ACM, 579.
\newblock


\bibitem{holstein2019improving}
{Kenneth Holstein}, {Jennifer Wortman~Vaughan}, {Hal Daum{\'e}~III}, {Miro
  Dudik}, {and} {Hanna Wallach}. 2019.
\newblock \showarticletitle{Improving fairness in machine learning systems:
  What do industry practitioners need?}. In {\em Proceedings of the 2019 CHI
  Conference on Human Factors in Computing Systems}. ACM, 600.
\newblock


\bibitem{johansson2009evolving}
{Ulf Johansson} {and} {Lars Niklasson}. 2009.
\newblock \showarticletitle{Evolving decision trees using oracle guides}. In
  {\em 2009 IEEE Symposium on Computational Intelligence and Data Mining}.
  IEEE, 238--244.
\newblock


\bibitem{kim2014bayesian}
{Been Kim}, {Cynthia Rudin}, {and} {Julie~A Shah}. 2014.
\newblock \showarticletitle{The bayesian case model: A generative approach for
  case-based reasoning and prototype classification}. In {\em Advances in
  Neural Information Processing Systems}. 1952--1960.
\newblock


\bibitem{kocielnik2019will}
{Rafal Kocielnik}, {Saleema Amershi}, {and} {Paul~N Bennett}. 2019.
\newblock \showarticletitle{Will You Accept an Imperfect AI?: Exploring Designs
  for Adjusting End-user Expectations of AI Systems}. In {\em Proceedings of
  the 2019 CHI Conference on Human Factors in Computing Systems}. ACM, 411.
\newblock


\bibitem{koh2017understanding}
{Pang~Wei Koh} {and} {Percy Liang}. 2017.
\newblock \showarticletitle{Understanding black-box predictions via influence
  functions}. In {\em Proceedings of the 34th International Conference on
  Machine Learning-Volume 70}. JMLR. org, 1885--1894.
\newblock


\bibitem{krause2016interacting}
{Josua Krause}, {Adam Perer}, {and} {Kenney Ng}. 2016.
\newblock \showarticletitle{Interacting with predictions: Visual inspection of
  black-box machine learning models}. In {\em Proceedings of the 2016 CHI
  Conference on Human Factors in Computing Systems}. ACM, 5686--5697.
\newblock


\bibitem{krishnan1999extracting}
{R Krishnan}, {G Sivakumar}, {and} {P Bhattacharya}. 1999.
\newblock \showarticletitle{Extracting decision trees from trained neural
  networks}.
\newblock {\em Pattern recognition\/} {32}, 12 (1999).
\newblock


\bibitem{kulesza2015principles}
{Todd Kulesza}, {Margaret Burnett}, {Weng-Keen Wong}, {and} {Simone Stumpf}.
  2015.
\newblock \showarticletitle{Principles of explanatory debugging to personalize
  interactive machine learning}. In {\em Proceedings of the 20th international
  conference on intelligent user interfaces}. ACM, 126--137.
\newblock


\bibitem{lai2018human}
{Vivian Lai} {and} {Chenhao Tan}. 2018.
\newblock \showarticletitle{On Human Predictions with Explanations and
  Predictions of Machine Learning Models: A Case Study on Deception Detection}.
\newblock {\em arXiv preprint arXiv:1811.07901\/} (2018).
\newblock


\bibitem{laugel2017inverse}
{Thibault Laugel}, {Marie-Jeanne Lesot}, {Christophe Marsala}, {Xavier Renard},
  {and} {Marcin Detyniecki}. 2017.
\newblock \showarticletitle{Inverse Classification for Comparison-based
  Interpretability in Machine Learning}.
\newblock {\em arXiv preprint arXiv:1712.08443\/} (2017).
\newblock


\bibitem{lim2009assessing}
{Brian~Y Lim} {and} {Anind~K Dey}. 2009.
\newblock \showarticletitle{Assessing demand for intelligibility in
  context-aware applications}. In {\em Proceedings of the 11th international
  conference on Ubiquitous computing}. ACM, 195--204.
\newblock


\bibitem{lim2010toolkit}
{Brian~Y Lim} {and} {Anind~K Dey}. 2010.
\newblock \showarticletitle{Toolkit to support intelligibility in context-aware
  applications}. In {\em Proceedings of the 12th ACM international conference
  on Ubiquitous computing}. ACM, 13--22.
\newblock


\bibitem{lim2009and}
{Brian~Y Lim}, {Anind~K Dey}, {and} {Daniel Avrahami}. 2009.
\newblock \showarticletitle{Why and why not explanations improve the
  intelligibility of context-aware intelligent systems}. In {\em Proceedings of
  the SIGCHI Conference on Human Factors in Computing Systems}. ACM,
  2119--2128.
\newblock


\bibitem{lipton2016mythos}
{Zachary~C Lipton}. 2016.
\newblock \showarticletitle{The mythos of model interpretability}.
\newblock {\em arXiv preprint arXiv:1606.03490\/} (2016).
\newblock


\bibitem{lou2013accurate}
{Yin Lou}, {Rich Caruana}, {Johannes Gehrke}, {and} {Giles Hooker}. 2013.
\newblock \showarticletitle{Accurate intelligible models with pairwise
  interactions}. In {\em Proceedings of the 19th ACM SIGKDD international
  conference on Knowledge discovery and data mining}. ACM, 623--631.
\newblock


\bibitem{lundberg2017unified}
{Scott~M Lundberg} {and} {Su-In Lee}. 2017.
\newblock \showarticletitle{A unified approach to interpreting model
  predictions}. In {\em Advances in Neural Information Processing Systems}.
  4765--4774.
\newblock


\bibitem{madumal2019grounded}
{Prashan Madumal}, {Tim Miller}, {Liz Sonenberg}, {and} {Frank Vetere}. 2019.
\newblock \showarticletitle{A Grounded Interaction Protocol for Explainable
  Artificial Intelligence}. In {\em Proceedings of the 18th International
  Conference on Autonomous Agents and MultiAgent Systems}. International
  Foundation for Autonomous Agents and Multiagent Systems, 1033--1041.
\newblock


\bibitem{miller2018explanation}
{Tim Miller}. 2018.
\newblock \showarticletitle{Explanation in artificial intelligence: Insights
  from the social sciences}.
\newblock {\em Artificial Intelligence\/} (2018).
\newblock


\bibitem{mohseni2018survey}
{Sina Mohseni}, {Niloofar Zarei}, {and} {Eric~D Ragan}. 2018.
\newblock \showarticletitle{A survey of evaluation methods and measures for
  interpretable machine learning}.
\newblock {\em arXiv preprint arXiv:1811.11839\/} (2018).
\newblock


\bibitem{molnar2018interpretable}
{Christoph Molnar} {and} {others}. 2018.
\newblock \showarticletitle{Interpretable machine learning: A guide for making
  black box models explainable}.
\newblock {\em E-book at< https://christophm. github.
  io/interpretable-ml-book/>, version dated\/}  {10} (2018).
\newblock


\bibitem{mothilal2020explaining}
{Ramaravind~K Mothilal}, {Amit Sharma}, {and} {Chenhao Tan}. 2020.
\newblock \showarticletitle{Explaining machine learning classifiers through
  diverse counterfactual explanations}. In {\em Proceedings of the 2020
  Conference on Fairness, Accountability, and Transparency}. 607--617.
\newblock


\bibitem{muller2019data}
{Michael Muller}, {Ingrid Lange}, {Dakuo Wang}, {David Piorkowski}, {Jason
  Tsay}, {Q~Vera Liao}, {Casey Dugan}, {and} {Thomas Erickson}. 2019.
\newblock \showarticletitle{How Data Science Workers Work with Data: Discovery,
  Capture, Curation, Design, Creation}. In {\em Proceedings of the 2019 CHI
  Conference on Human Factors in Computing Systems}. ACM, 126.
\newblock


\bibitem{narayanan2018humans}
{Menaka Narayanan}, {Emily Chen}, {Jeffrey He}, {Been Kim}, {Sam Gershman},
  {and} {Finale Doshi-Velez}. 2018.
\newblock \showarticletitle{How do humans understand explanations from machine
  learning systems? an evaluation of the human-interpretability of
  explanation}.
\newblock {\em arXiv preprint arXiv:1802.00682\/} (2018).
\newblock


\bibitem{nguyen2016multifaceted}
{Anh Nguyen}, {Jason Yosinski}, {and} {Jeff Clune}. 2016.
\newblock \showarticletitle{Multifaceted feature visualization: Uncovering the
  different types of features learned by each neuron in deep neural networks}.
\newblock {\em arXiv preprint arXiv:1602.03616\/} (2016).
\newblock


\bibitem{poursabzi2018manipulating}
{Forough Poursabzi-Sangdeh}, {Daniel~G Goldstein}, {Jake~M Hofman},
  {Jennifer~Wortman Vaughan}, {and} {Hanna Wallach}. 2018.
\newblock \showarticletitle{Manipulating and measuring model interpretability}.
\newblock {\em arXiv preprint arXiv:1802.07810\/} (2018).
\newblock


\bibitem{rader2018explanations}
{Emilee Rader}, {Kelley Cotter}, {and} {Janghee Cho}. 2018.
\newblock \showarticletitle{Explanations as mechanisms for supporting
  algorithmic transparency}. In {\em Proceedings of the 2018 CHI Conference on
  Human Factors in Computing Systems}. ACM, 103.
\newblock


\bibitem{ram1989question}
{Ashwin Ram}. 1989.
\newblock \showarticletitle{Question-driven understanding: An integrated theory
  of story understanding, memory and learning}.
\newblock  (1989).
\newblock


\bibitem{ras2018explanation}
{Gabri{\"e}lle Ras}, {Marcel van Gerven}, {and} {Pim Haselager}. 2018.
\newblock \showarticletitle{Explanation methods in deep learning: Users,
  values, concerns and challenges}.
\newblock In {\em Explainable and Interpretable Models in Computer Vision and
  Machine Learning}. Springer, 19--36.
\newblock


\bibitem{ribeiro2016should}
{Marco~Tulio Ribeiro}, {Sameer Singh}, {and} {Carlos Guestrin}. 2016.
\newblock \showarticletitle{Why should i trust you?: Explaining the predictions
  of any classifier}. In {\em Proceedings of the 22nd ACM SIGKDD international
  conference on knowledge discovery and data mining}. ACM, 1135--1144.
\newblock


\bibitem{ribeiro2018anchors}
{Marco~Tulio Ribeiro}, {Sameer Singh}, {and} {Carlos Guestrin}. 2018.
\newblock \showarticletitle{Anchors: High-precision model-agnostic
  explanations}. In {\em Thirty-Second AAAI Conference on Artificial
  Intelligence}.
\newblock


\bibitem{ribes2017notes}
{David Ribes}. 2017.
\newblock \showarticletitle{Notes on the concept of data interoperability:
  Cases from an ecology of AIDS research infrastructures}. In {\em Proceedings
  of the 2017 ACM Conference on Computer Supported Cooperative Work and Social
  Computing}. ACM, 1514--1526.
\newblock


\bibitem{robnik2018perturbation}
{Marko Robnik-{\v{S}}ikonja} {and} {Marko Bohanec}. 2018.
\newblock \showarticletitle{Perturbation-Based Explanations of Prediction
  Models}.
\newblock In {\em Human and Machine Learning}. Springer, 159--175.
\newblock


\bibitem{rule2018exploration}
{Adam Rule}, {Aur{\'e}lien Tabard}, {and} {James~D Hollan}. 2018.
\newblock \showarticletitle{Exploration and explanation in computational
  notebooks}. In {\em Proceedings of the 2018 CHI Conference on Human Factors
  in Computing Systems}. ACM, 32.
\newblock


\bibitem{samek2019towards}
{Wojciech Samek} {and} {Klaus-Robert M{\"u}ller}. 2019.
\newblock \showarticletitle{Towards explainable artificial intelligence}.
\newblock In {\em Explainable AI: Interpreting, Explaining and Visualizing Deep
  Learning}. Springer, 5--22.
\newblock


\bibitem{sandvig2014auditing}
{Christian Sandvig}, {Kevin Hamilton}, {Karrie Karahalios}, {and} {Cedric
  Langbort}. 2014.
\newblock \showarticletitle{Auditing algorithms: Research methods for detecting
  discrimination on internet platforms}.
\newblock {\em Data and discrimination: converting critical concerns into
  productive inquiry\/}  {22} (2014).
\newblock


\bibitem{schneider2019personalized}
{Johanes Schneider} {and} {Joshua Handali}. 2019.
\newblock \showarticletitle{Personalized explanation in machine learning}.
\newblock {\em arXiv preprint arXiv:1901.00770\/} (2019).
\newblock


\bibitem{silveira2001semiotic}
{Milene~Selbach Silveira}, {Clarisse~Sieckenius de Souza}, {and} {Simone~DJ
  Barbosa}. 2001.
\newblock \showarticletitle{Semiotic engineering contributions for designing
  online help systems}. In {\em Proceedings of the 19th annual international
  conference on Computer documentation}. ACM, 31--38.
\newblock


\bibitem{simonyan2013deep}
{Karen Simonyan}, {Andrea Vedaldi}, {and} {Andrew Zisserman}. 2013.
\newblock \showarticletitle{Deep inside convolutional networks: Visualising
  image classification models and saliency maps}.
\newblock {\em arXiv preprint arXiv:1312.6034\/} (2013).
\newblock


\bibitem{springer2019progressive}
{Aaron Springer} {and} {Steve Whittaker}. 2019.
\newblock \showarticletitle{Progressive disclosure: empirically motivated
  approaches to designing effective transparency}. In {\em Proceedings of the
  24th International Conference on Intelligent User Interfaces}. ACM, 107--120.
\newblock


\bibitem{vstrumbelj2014explaining}
{Erik {\v{S}}trumbelj} {and} {Igor Kononenko}. 2014.
\newblock \showarticletitle{Explaining prediction models and individual
  predictions with feature contributions}.
\newblock {\em Knowledge and information systems\/} {41}, 3 (2014), 647--665.
\newblock


\bibitem{stumpf2007toward}
{Simone Stumpf}, {Vidya Rajaram}, {Lida Li}, {Margaret Burnett}, {Thomas
  Dietterich}, {Erin Sullivan}, {Russell Drummond}, {and} {Jonathan Herlocker}.
  2007.
\newblock \showarticletitle{Toward harnessing user feedback for machine
  learning}. In {\em Proceedings of the 12th international conference on
  Intelligent user interfaces}. ACM, 82--91.
\newblock


\bibitem{swartout1983xplain}
{William~R Swartout}. 1983.
\newblock \showarticletitle{XPLAIN: A system for creating and explaining expert
  consulting programs}.
\newblock {\em Artificial intelligence\/} {21}, 3 (1983), 285--325.
\newblock


\bibitem{swartout1987making}
{William~R Swartout} {and} {Stephen~W Smoliar}. 1987.
\newblock \showarticletitle{On making expert systems more like experts}.
\newblock {\em Expert Systems\/} {4}, 3 (1987), 196--208.
\newblock


\bibitem{thomaz2006transparency}
{Andrea~L Thomaz} {and} {Cynthia Breazeal}. 2006.
\newblock \showarticletitle{Transparency and socially guided machine learning}.
  In {\em 5th Intl. Conf. on Development and Learning (ICDL)}.
\newblock


\bibitem{tolomei2017interpretable}
{Gabriele Tolomei}, {Fabrizio Silvestri}, {Andrew Haines}, {and} {Mounia
  Lalmas}. 2017.
\newblock \showarticletitle{Interpretable predictions of tree-based ensembles
  via actionable feature tweaking}. In {\em Proceedings of the 23rd ACM SIGKDD
  international conference on knowledge discovery and data mining}. ACM,
  465--474.
\newblock


\bibitem{wachter2017counterfactual}
{Sandra Wachter}, {Brent Mittelstadt}, {and} {Chris Russell}. 2017.
\newblock \showarticletitle{Counterfactual Explanations without Opening the
  Black Box: Automated Decisions and the GPDR}.
\newblock {\em Harv. JL \& Tech.\/}  {31} (2017), 841.
\newblock


\bibitem{wang2019designing}
{Danding Wang}, {Qian Yang}, {Ashraf Abdul}, {and} {Brian~Y Lim}. 2019.
\newblock \showarticletitle{Designing Theory-Driven User-Centric Explainable
  AI}. In {\em Proceedings of the 2019 CHI Conference on Human Factors in
  Computing Systems}. ACM, 601.
\newblock


\bibitem{wei2019generalized}
{Dennis Wei}, {Sanjeeb Dash}, {Tian Gao}, {and} {Oktay Gunluk}. 2019.
\newblock \showarticletitle{Generalized Linear Rule Models}. In {\em
  International Conference on Machine Learning}. 6687--6696.
\newblock


\bibitem{weld2018challenge}
{Daniel~S Weld} {and} {Gagan Bansal}. 2018.
\newblock \showarticletitle{The challenge of crafting intelligible
  intelligence}.
\newblock {\em arXiv preprint arXiv:1803.04263\/} (2018).
\newblock


\bibitem{weller2017challenges}
{Adrian Weller}. 2017.
\newblock \showarticletitle{Challenges for transparency}.
\newblock {\em arXiv preprint arXiv:1708.01870\/} (2017).
\newblock


\bibitem{wolf2019explainability}
{Christine~T Wolf}. 2019.
\newblock \showarticletitle{Explainability scenarios: towards scenario-based
  XAI design}. In {\em Proceedings of the 24th International Conference on
  Intelligent User Interfaces}. ACM, 252--257.
\newblock


\bibitem{yang2018machine}
{Qian Yang}. 2018.
\newblock \showarticletitle{Machine Learning as a UX Design Material: How Can
  We Imagine Beyond Automation, Recommenders, and Reminders?}. In {\em 2018
  AAAI Spring Symposium Series}.
\newblock


\bibitem{yin2019understanding}
{Ming Yin}, {Jennifer Wortman~Vaughan}, {and} {Hanna Wallach}. 2019.
\newblock \showarticletitle{Understanding the Effect of Accuracy on Trust in
  Machine Learning Models}. In {\em Proceedings of the 2019 CHI Conference on
  Human Factors in Computing Systems}. ACM, 279.
\newblock


\bibitem{zhang2019interpreting}
{Quanshi Zhang}, {Yu Yang}, {Haotian Ma}, {and} {Ying~Nian Wu}. 2019.
\newblock \showarticletitle{Interpreting cnns via decision trees}. In {\em
  Proceedings of the IEEE Conference on Computer Vision and Pattern
  Recognition}. 6261--6270.
\newblock


\bibitem{zhang2018interpreting}
{Xin Zhang}, {Armando Solar-Lezama}, {and} {Rishabh Singh}. 2018.
\newblock \showarticletitle{Interpreting neural network judgments via minimal,
  stable, and symbolic corrections}. In {\em Advances in Neural Information
  Processing Systems}. 4874--4885.
\newblock


\bibitem{zhou2016learning}
{Bolei Zhou}, {Aditya Khosla}, {Agata Lapedriza}, {Aude Oliva}, {and} {Antonio
  Torralba}. 2016.
\newblock \showarticletitle{Learning deep features for discriminative
  localization}. In {\em Proceedings of the IEEE conference on computer vision
  and pattern recognition}. 2921--2929.
\newblock


\bibitem{zhou2003extracting}
{Zhi-Hua Zhou}, {Yuan Jiang}, {and} {Shi-Fu Chen}. 2003.
\newblock \showarticletitle{Extracting symbolic rules from trained neural
  network ensembles}.
\newblock {\em Ai Communications\/} {16}, 1 (2003), 3--15.
\newblock


\bibitem{zhu2018explainable}
{Jichen Zhu}, {Antonios Liapis}, {Sebastian Risi}, {Rafael Bidarra}, {and}
  {G~Michael Youngblood}. 2018.
\newblock \showarticletitle{Explainable AI for designers: A human-centered
  perspective on mixed-initiative co-creation}. In {\em 2018 IEEE Conference on
  Computational Intelligence and Games (CIG)}. IEEE, 1--8.
\newblock


\end{thebibliography}

\end{document}